\documentclass[twocolumn]{aastex63}
\usepackage{xcolor}
\usepackage{amsmath}
\usepackage{amstext}
\usepackage{apjfonts}
\usepackage{bm}
\usepackage{multirow}
\usepackage{hyperref}
\usepackage[figure,figure*]{hypcap}
\usepackage[graphicx]{realboxes}
\usepackage{xurl}
\usepackage{enumitem}
\usepackage[normalem]{ulem}
\graphicspath{{figures/}}

\begin{document}

\title{Attaining Spectral Energy Distributions With Sub-Percent Uncertainties:  All-Sky DA White Dwarf Spectrophotometric Standard Stars For Large Telescopes And Surveys.}

\author[0000-0002-6839-4881]{Abhijit Saha}
\affiliation{NSF's National Optical Infrared Astronomy Research Laboratory, 950 North Cherry Avenue, Tucson, AZ 85719}
\author[0000-0002-7157-500X]{Edward W. Olszewski}
\affiliation{The University of Arizona, Steward Observatory, 933 North Cherry Avenue, Tucson, AZ 85719}
\author[0000-0002-0622-1117]{Benjamin M. Boyd}
\affiliation{Institute of Astronomy and Kavli Institute for Cosmology, University of Cambridge, Madingley Road, Cambridge, CB3 0HA, UK}
\author[0000-0001-6685-0479]{Thomas Matheson}
\affiliation{NSF's National Optical Infrared Astronomy Research Laboratory, 950 North Cherry Avenue, Tucson, AZ 85719}
\author [0000-0002-5722-7199]{Tim Axelrod}
\affiliation{The University of Arizona, Steward Observatory, 933 North Cherry Avenue, Tucson, AZ 85719}
\author[0000-0001-6022-0484]{Gautham Narayan}
\affiliation{University of Illinois at Urbana-Champaign, 1002 W. Green St., Urbana IL 61801}
\author[0000-0002-0882-7702]{Annalisa Calamida}
\affiliation{Space Telescope Science Institute, 3700 San Martin Drive, Baltimore, MD 21218}
\affiliation{INAF - Osservatorio Astronomico Capodimonte, Salita Moiariello, 16, 80131 Napoli, Italy
 }
\author [0000-0003-3082-0774]{Jay B. Holberg}
\affiliation{The University of Arizona, Lunar and Planetary Laboratory, 1629 East University Boulevard, Tucson, AZ 85721}
\author [0000-0001-8816-236X]{Ivan Hubeny}
\affiliation{The University of Arizona, Steward Observatory, 933 North Cherry Avenue, Tucson, AZ 85719}
\author[0000-0001-9806-0551]{Ralph C. Bohlin}
\affiliation{Space Telescope Science Institute, 3700 San Martin Drive, Baltimore, MD 21218}
\author[0000-0003-2823-360X]{Susana Deustua}
\affiliation{Sensor Science Division, National Institute of Standards and Technology, Gaithersburg, MD 20899-8441}
\affiliation{Space Telescope Science Institute, 3700 San Martin Drive, Baltimore, MD 21218}
\author[0000-0002-4410-5387]{Armin Rest}
\affiliation{Space Telescope Science Institute, 3700 San Martin Drive, Baltimore, MD 21218}
\affiliation{Department of Physics and Astronomy, Johns Hopkins University, Baltimore, MD 21218, USA}
\author{Jenna Claver}
\affiliation{NSF's National Optical Infrared Astronomy Research Laboratory, 950 North Cherry Avenue, Tucson, AZ 85719}
\author [0000-0002-4596-1337]{Sean Points}
\affiliation{NSF's National Optical Infrared Astronomy Research Laboratory, 950 North Cherry Avenue, Tucson, AZ 85719}
\author[0000-0003-0347-1724]{Christopher W. Stubbs}
\affiliation{Harvard University, Department of Physics, 17 Oxford Street, Cambridge, MA 02138}
\affiliation{Harvard-Smithsonian Center for Astrophysics, 60 Garden Street, Cambridge, MA 02138}
\author [0000-0003-2954-7643]{Elena Sabbi}
\affiliation{NSF's National Optical Infrared Astronomy Research Laboratory, 950 North Cherry Avenue, Tucson, AZ 85719}
\affiliation{Space Telescope Science Institute, 3700 San Martin Drive, Baltimore, MD 21218}
\author[0000-0001-6529-8416]{John W. Mackenty}
\affiliation{Space Telescope Science Institute, 3700 San Martin Drive, Baltimore, MD 21218}

\begin{abstract}
    We present a synopsis of the project to establish thirty-two new faint ($ 16.5 \leq V \leq 19.8 $) DA white dwarfs as spectrophotometric standards distributed over the whole sky. 
    Our results validate the use of fully radiative pure hydrogen model fluxes for hot DA white dwarfs to predict the observed 
    broadband fluxes from near ultraviolet through the near infrared to accuracies of a few parts per thousand. After fitting the line of sight reddenings simultaneously with the model spectral energy distributions of these stars against spectroscopic and multi-band photometric observations, we have shown that residuals have an rms of typically 0.4 percent.
    This indicates that the complications from interstellar dust extinction have been adequately mitigated.
    Our stars supplement the three brighter DA white dwarfs that define the flux scale of CALSPEC.  The consequent photometric accuracy,
    their all sky coverage, and their brightness range that matches the dynamic range of large telescopes, constitutes an unprecedented ensemble of standard stars for both ground as well as space based use.
     This paper targets readers who may wish to use these as standard stars, and provides for them the essential content to understand their strengths and limitations, without traversing the technical details of analysis that are already captured in a series of papers since 2016. 
     The narrative here describes 
     the motivation, justification, and evolution of the analysis methods; the input data that constrain the modeling; as well as the stability of our results in the face of future improvements in models. 
\end{abstract}


\section{Introduction}
\label{sec:Intro}

In a decade-long series of papers, \citet{Narayan16}, \citet{Calamida19}, \citet{Narayan19}, \citet{Calamida22}, \citet{axelrod2023},  and culminating in \citet{Boyd2025},
we have documented the observations, methodology and results for selecting, testing, and calibrating a set of thirty five DA white dwarf (DAWDs) as spectrophotometric standard stars. To the three relatively bright DA white dwarfs that have anchored the CALSPEC fluxes for over two decades, we have added thirty-two DAWDs in the magnitude range $16.5 \le V \le 19.8$ (with the majority around $V \approx 18$), which places them in the optimal signal-to-noise ratio dynamic range for observations with the largest telescopes and space based facilities in use today.  The full ensemble, including the three bright DAWDs is thus 35 stars.

In the spectral range from $\approx 275$ nm  in the ultraviolet (UV) through $\approx 1600$ nm (H-band) in the near infrared (IR), our ensemble of DAWDs are validated to be accurate for broadband photometry at the 
{uncertainty level of a} few $\times$ 0.1\%, when compared against
space based broadband photometry 
obtained with the Wide Field Camera 3 (WFC3) instrument on board the Hubble Space Telescope (HST) using 5 filters in the UV/Visible (UVIS) channel ($F275W, F336W, F475W, F625W, F775W$) and the $F160W$ filter in the near-infrared (IR) channel in 3 HST programs in Cycle 20 (Program 12967; 2012-2013), Cycle 22 (Program 13711; 2014-2016), and Cycle 25 (Program 15113; 2018).
 
To our knowledge, this network of 35 stars constitutes one of the most accurate and coherently calibrated set of photometric standards.  
They are available over the entire sky and 
the thirty-two
fainter stars present a “sweet spot” in terms of dynamic range and signal-to-noise, accessible to the largest aperture telescopes and modern space facilities.

The co-analysis of HST Space Telescope Imaging Spectrograph (STIS) UV spectra of 19 of these stars described by \citet{Boyd2025} leave the model fluxes in the near-UV through near-IR wavelengths essentially unchanged, which observationally validates the models as standard fluxes further into the UV to
$\approx$ 130 nm for the subset of our ensemble with available STIS spectra.  Full confidence in this conclusion awaits the analysis and successful validation of \emph{all} of our standard stars. For the UV and IR calibrations to attain the same reliability as that achieved here for the spectral coverage of WFC3, will require data in the UV and in the IR with the same demonstrable internal consistency across all objects, along with showing that the synthetic fluxes and interstellar extinction behavior continue to reproduce the observed fluxes at those outlying wavelengths. 


Subsets of our standard stars are embedded in the survey areas of existing and future programs such as the Panoramic Survey Telescope \& Rapid Response System (PanSTARRS), the Sloan Digital Sky Survey (SDSS), the Dark Energy Survey (DES), Gaia, and Euclid, and 
in forthcoming future surveys, notably the NSF-DOE Vera C. Rubin Observatory's Legacy Survey of Space and Time (LSST), the Nancy Grace Roman Space Telescope (Roman), and NASA's projected Habitable Worlds Observatory.

The above cited papers, and the results presented therein, cover all the relevant details of this decade-long effort. Anyone interested in the details of how the experiment was conducted as well as how the data were processed and analyzed (in various stages) can find   full descriptions there. However, we recognize that many potential users of the standard stars will be interested in the strengths, caveats, and limitations of these DAWD standards, including the underlying assumptions inherent in defining them as spectrophotometric standard stars.  The purpose of this paper is to lay out this information so that a potential user of these standard stars is able to decide whether they are suitable for the purpose they have in mind, without having to delve into the full details of the methods and analysis employed.  We have made a conscious decision not to repeat figures and tables from earlier papers, but to cite them at their original locations. 

\subsection{Why DA white dwarfs?}
\label{sec:whyDAWDs}
The over-arching rationale for choosing DAWDs is that, by definition, these stars have pure hydrogen atmospheres. Moreover, hot ($T > 15000 \thinspace K$) DAWDs are fully radiative and not subject to pulsational instabilities. Their photospheres are completely characterized by two parameters: the effective temperature, $T_{e}$, and the surface gravity $\log \thinspace g$ \citep[e.g.,][]{Holbergetal1985}.  This means that their atmospheres are stable and therefore not intrinsically variable in brightness.  Being of pure hydrogen, they are the least complicated to model of any class of stellar atmospheres.  Thus, of all the various kinds of stars, these appear to be the ideal candidates for which one can expect the best predicted spectral energy distributions (SEDs) from modeling.  \citet{Bohlin_2000} demonstrated, with three DAWDs, that their spectroscopically determined respective values of $T_{e}$ and $\log \thinspace g$ predicted their relative SEDs to better than $1 \%$. The CALSPEC system, which provides flux standards for HST, and the James Webb Space Telescope (JWST), is anchored on the model-predicted SEDs of these three
stars \citep{Bohlin_2020}.  

Our experiment compares the detected flux above the terrestrial atmosphere using WFC3 on HST to the predicted SED as informed by ground-based spectra as described through the chain of papers cited above. We have sought to test just how well
the model atmospheres predict the SEDs for a larger set of 
35 DAWDs:   32  faint DAWDs plus the three DAWDs that anchor the CALSPEC system.
As shown by \citet[hereafter A23]{axelrod2023}, within the wavelength range from $ \approx 270 {\rm ~nm ~to~} \approx 900 {\rm ~nm} $, non-local thermodynamical equilibrium (NLTE) white dwarf atmosphere models predict the wavelength dependence of the SEDs for these 35 stars to within an rms 
uncertainty
of $\approx 0.004$ mag 
for broadband passbands. At $1600 {\rm ~nm}$, where the IR channel of WFC3 is used, the count-rate non-linearity\footnote{https://hst-docs.stsci.edu/wfc3dhb/chapter-7-wfc3-ir-sources-of-error/7-7-count-rate-non-linearity} must be accounted for. The error weighted rms for the sample  
 is $\approx 0.006$ mag, but is biased by the brighter objects of the ensemble.
 The unweighted rms at $1600 {\rm ~nm}$ is $\approx 0.008$ mag and includes the three bright DAWDs. 

\citet{Boyd2025} (hereafter B25) represents the final analysis with the fully matured methodology, which uses updated DAWD model atmospheres and dust models for line-of-sight reddening, joint inference methods for both photometry and spectra to mitigate correlated errors, and incorporates modeling and comparison of observed STIS UV spectra where available (for 19 of the 35 DAWDs). This method produces a reduction of the rms residuals by up to 0.001 mag compared to those in A23. The results illustrate the robustness of the analysis to changes in the models and to extending the wavelength range, as described further in \S~\ref{sec:anal-meth} and \S~\ref{sec:Comps}.  
For calibrating existing optical surveys,  our synthetic magnitudes for those surveys based on our derived SEDs (B25) and on published survey passbands are available on Zenodo (see \S~\ref{sec:results})
 
In \S~\ref{sec:needper} we describe the motivation for our work, and the factors that guided our experimental design, while \S~\ref{sec:history} follows the evolution of both scope and methodology of our work and the reasons for them.  The intercomparison of results from various stages of our work, as well as with results from alternative analyses (outside the confines of our methodology, but using data for this project) provides perspective about overall robustness and dependencies on models. This is covered in \S~\ref{sec:Comps}. Future efforts that can test and expand the reach of these standards to other wavelengths (e.g., making them useful for JWST, and establishing standardization in the UV) is under development by our team and by others, with a general discussion and summary in \S~\ref{sec:discussion}.

\section{Needs and Perspectives}
\label{sec:needper}

There has been a pressing need for a way to tie spectrophotometric measurements over the whole sky in absolute physical units linked to the SI International System of units 
(we use the cgs alternative, and the related AB magnitudes familiar to astronomers) to sub-percent uncertainties.
Cosmological studies, such as those using faint Type Ia Supernovae as probes, or those relying on photometric redshifts (photo-z)
require uncertainties smaller 
than a percent in colors (i.e. flux ratios). Specifically, the uncertainty in determining the equation of state for dark energy is dominated by 
measurement uncertainties in the photometry, as exemplified in \citet{Betoule2014} and \citet{Hounsell2018}. System-wide 
consistency in calibration across ground and space observations, across wavelength ranges probed by different instrumentation, as well as assurance of linearity across 
all brightness levels of interest to astronomy depend on the mitigation of systematic errors in calibration.


An astronomical point source illuminates a telescope's pupil with a collimated beam. The telescope focuses such an object onto a detector as a point-like image, whose intensity can be recorded and measured (after adjusting, as best as possible, for instrumental signatures, such as bias corrections and flat fields). The problem at hand is to translate that measurement onto an absolute scale with physical units, i.e., in terms of, say, $\rm ergs \thinspace cm^{-2} s^{-1} Hz^{-1}$, and infer what that flux would be if received above the terrestrial atmosphere.  There is 
a rich history of attempts to do so beginning with the pioneering work of \citet{Oke1970}, \citet{Hayes1970}, \citet{HayesLath1975}, and  \citet{OkeGunn83}. The seminal work of setting up the CALSPEC standards (e.g. \citet{Bohlin_2014}) based on the SEDs of 3 bright (V$\sim 12$ mag) DAWDs has been the state of the art, with percent-level uncertainties relative to the models.
However, a set of standard stars faint enough for modern day large-aperture telescopes with sub-percent systemic uncertainties
has been elusive.  Our work has been motivated by the need to address this shortcoming.


SI-traceable spectrophotometric measurements at specific wavelengths exist for Vega in the visible (cf. the above cited papers).   \citet{Megessier95}'s analysis of the published results of a half dozen spectroscopic experiments provided a value of the flux at $555.6$ nm for Vega. \citet{Price2004} provides photometric fluxes for Sirius and Vega at wavelengths beyond 4 $\mu$m (microns). Vega is known to have circumstellar dust \citep{Su2005}, and is incompatible with Sirius at the mid-IR wavelengths where the latter has accurate absolute flux determinations.  Thus, at this time, independent experimental validation of models for Vega and Sirius remain uncertain at the order of $1 \%$, but the absolute flux at certain specific wavelengths for Vega and for Sirius are known.  In the absence of high accuracy direct measurements based on laboratory SI standards, absolute stellar fluxes have been inferred from modeling the observed features of certain types of stellar atmospheres. The modeling uses the resulting inferred physical parameters of the stellar photosphere to  produce an SED which is normalized to bright stars such as Vega and Sirius.  We are unaware of any \emph{independent} observational validation of the SEDs spanning large wavelength baselines (UV through mid-infrared, panchromatic) of such models to better than $\sim 1\%$. 



We recognize that the problem can be partitioned into two independent parts:
\begin{enumerate}
    \item The wavelength dependence of the SED, which need only be measured in a relative way for the same object at different wavelengths. This is sometimes referred to as the chromatic or color component. 
    
    \item The absolute received flux  for at least one wavelength that is ultimately tied to absolute physical units. 
\end{enumerate}
These two components together provide a complete specification, but, in practice, present very different complexities when we try to measure them. SEDs can be measured as the flux ratios at different wavelengths for an unvarying source, and, for a stable detector, can be calibrated against sources with known SEDs, such as standardized laboratory sources. Absolute brightness, however, requires knowledge of the actual rate of incidence of photons at the top of the atmosphere, which is beset by its own set of additional complications.  \\

Our experiments have shown that the wavelength dependence of SEDs (hereafter wSED) can be determined for astronomical sources to 
very high accuracies, achieving sub-percent uncertainties,
whereas the achromatic normalization of the SED (nSED) leaves room for improvement, and is currently uncertain for astronomical applications at the $ 1 \%~ {\rm to } ~2 \% $  (1-sigma) level.  Astrophysical applications, such as characterization of the host stars of exo-planets, will benefit from improvements in the accuracy of the absolute achromatic zero-point \citep[e.g.][]{Tayar_2022}, whereas sub-percent calibrations of "colors" is, for example, the dominant term in the error budget for determining the dark-energy equation of state using galaxy photo-z's.
Our methods bring significant improvements to the accuracy of wSEDs, but   rely on external constraints for nSED. The disambiguation of these two components in the calibration will recur in the discussion of our results later in this paper. 
\\
\subsection{Instrumental vs. Atmospheric Modulation}
\label{INSTvsATM}
Consider first a ground-based facility, with a pixelated imaging device at the focal plane of a telescope.
The detected time integrated signal in counts (ADU) from incident flux $F_{\lambda}$ at wavelength $\lambda$ detected from a source at a location $(x,y)$ on the detector can be written as:
\begin{equation}
    ADU ~~=~~ A~\int_{0}^{\Delta T} dt \int_{0}^{\infty} \lambda^{-1}~ F_{\lambda} ~ S^{tot}(x, y, alt, az, t, \lambda)~ d {\lambda} 
\end{equation}
where $S^{tot}$ represents the system response that, in general, depends on the position on the detector ($x, y$), the direction the telescope is pointed ($alt, az$, which accounts for atmospheric extinction), and temporal and wavelength dependent variations in the response ($t$, and $\lambda$).  The factor $A$ is a constant for the system, and includes the collecting area of the telescope and other invariants.
\\
It is reasonable to write $S^{tot}$ as being composed of two mutually independent parts:
\begin{equation}
   S^{tot}(x, y, alt, az, t, \lambda) = S^{inst}(x, y, t, \lambda) \times S^{atm}( {\rm alt, az}, t, \lambda)
\end{equation}
where $ S^{inst} $ captures the response from the top of the telescope through to the detector and $S^{atm} $ accounts for the modulation suffered through the terrestrial atmosphere, which can depend on direction and time, as well as the wavelength.  For a telescope in space, $S^{atm}$ is effectively unity and invariant. \\

It is possible to determine and monitor $S^{inst}$ in "laboratory" mode, e.g., with techniques described by \citet{Woodward2010}  and demonstrated by the DECal Spectrophotometric calibration system \citet{Marshall_2016} for calibrating the instrumental response of the Dark Energy Camera (DECam). Such a setup 
is specifically built into the Rubin observatory infrastructure, where flat fields illuminated by tuneable (wavelength) lasers can be used to map the "monochromatic" response at all wavelengths across the face of the detector \citep{Ingraham_2022}.  Care must be exercised to ascertain that the telescope pupil is uniformly illuminated, with a well-understood angular dependence, so that the system response for 
point sources as well as the surface brightness of extended sources can be quantified. Absolute energies entering the telescope pupil can be measured with {\it NIST}\footnote{National Institute of Standards and Technology} detectors calibrated to exacting standards.  It would seem difficult and perhaps prohibitive to outfit telescopes in space in the same way.  A much-dreamed option is to "fly" a standard source in space (or very high above most of the terrestrial atmosphere in a balloon) that can be accessed both from the ground and from space-based observatories. To date, such facilities have not come to fruition. At the time of this writing, several space- and balloon-borne experiments are under development.

Direct determination and/or measurement informed modeling of $S^{atm}$ is much more complex, but efforts are underway to measure and monitor the atmospheric extinction and other telluric effects in real time. For progress on this front, see for example: \citet{Pederson2025,Smette2020SPIE11449E..1HS}.
 
Using real stars as calibrators has the practical advantage that they are observed in the same manner as other "unknown" astronomical targets.  This finesses the problem of ascertaining that the calibration source illuminates the telescope in the same way as the unknowns.  Incidentally, this is also the argument for having standard stars within the dynamic range of the telescope and instrument system, as practices such as de-focusing the telescope to accommodate bright standards by spreading their light on the detector affects the illumination pattern through the optics and can introduce systematic differences in $S^{tot}$. The earliest experiments to bring astronomical flux measurements to an absolute scale calibrated the bright star Vega against physical laboratory standards \citep{Oke1970, Hayes1970}, giving birth to spectrophotometry. The goal then, as now, is to set select stellar objects onto the physical absolute flux scale and to subsequently use the star(s) as standards.  With our growth in the understanding of stellar atmospheres and our ability to model them, it soon became apparent that, for select kinds of stars, such models produced better predictions of the SEDs of such standard stars than the direct calibration from ground-based laboratory sources. \citet{Bohlin-Gilliland2004} showed this in a definitive manner. It is safe to state that the most self-consistent calibrators in use today are from models of stellar atmospheres. 

However, we do not yet have the final word on nSED, the achromatic normalization. The fidelity
of Vega, long used as a primary standard star, is questionable at the few percent level, since we now know that it has dust contamination around it (\citet{Aumann1984}, \citet{Su2005} \citet{Sibthorpe2010}). We are seeing progress in developing experiments by other teams who are using direct calibration methods with satellite and balloon experiments \citep{Plav2023, Deustua2021}.  

We claim that our experiment succeeds in bringing down the uncertainties in wSED to $\sim 0.4 \%$ (1-sigma) in the visible, and  to $\sim 0.7\%$ (1-sigma) in the near-IR.  

To proceed using models of stellar atmospheres as predictors of wSEDs, we must choose wisely.  With this in mind, and for reasons detailed below, we embarked on testing the efficacy of using the predicted SEDs of hot DAWDs from their atmospheric models.  The various stages of our project are recounted by A23, which traces the evolution of our methods through \citet[hereafter N16]{Narayan16}, \citet[hereafter C19]{Calamida19}, \citet[hereafter N19]{Narayan19}, and \citet[hereafter C22]{Calamida22}. 
\subsection{Methodology}
\label{sec:methodology}
Keeping the above discussion in mind, we set our focus on testing the predictability of wavelength dependence of the SEDs, i.e, the wSED component, from DAWD models to greater accuracy. However, we must recognize that re-examining the achromatic absolute flux zero-point would require a very different experiment. At the time when we began our investigation, the demand clearly was for improving wSED, since it affects a variety of issues:
\begin{enumerate}[label=(\alph*)]
\item  Accuracy in wSED limits the accuracy for photometric redshift determinations.     
\item  The availability of accurate wSEDs allows reliably accurate transformations across different photometric systems, if the transmission functions of the passbands are known.
\item  Most physical processes under investigation require accurate wSED to sub-percent uncertainty levels. In contrast, a 1\% or 2\% uncertainty in setting an nSED zero-point is not as critical. This has changed since the time we designed our experiment, as, for instance, exoplanet work can demand sub-percent accuracies in nSED as well. 
\end{enumerate}

Our experiment has three observational components: 

\begin{enumerate}
    \item Ground-based spectra from which we can derive the physical parameters $T_{e}$ and $\log g$, which in turn predicts the SED emerging from the DAWD star.  A variety of facilities were used to obtain these spectra as described further below.  At a late stage of our effort, UV spectra of 19 of the faint DAWD stars were obtained with HST/STIS, to help test the validity of our inferred extinction/reddening, as described by B25.
    
    
    \item Broadband photometry (over a wide wavelength range) of potential candidate DAWDs, obtained above the atmosphere to eliminate atmospheric effects, referred to an instrumental system that is internally standardized across all observations.
    
    \item 
    Ground-based monitoring of all candidates for variability on a variety of time-scales over a period of two years, to demonstrate that they are indeed non-variable.
\end{enumerate}

The evolution of how we have treated 1) above is described in \S\ref{sec:history}.  For item 2) above, internal standardization was achieved for the photometry with WFC3 on board the 
HST, by contemporaneous measurement of the 3 
primary CALSPEC DAWD standard stars from \citet{Bohlin07}.
These 3 stars 
can trace the sensitivity of the instrument over time spans when the remaining candidate DAWDs were observed. 
For item 3) we monitored the brightness of our target stars against neighboring objects in the field, using the Las Cumbres Observatory (LCO) Network of Telescopes. We sampled over a variety of time scales, over a total duration of $\sim 2$ years. This allowed us to ensure that our ensemble is free of objects showing any variability, such as from debris disks.  
For those interested, details of all observational aspects listed above can be found in C19 and C22.

The notional procedure is to compare the SEDs predicted by the spectra (which give us $T_{e}$ and $\log g$, the two parameters that completely specify the atmospheric model for DAWDs) to the internally standardized instrumental pan-chromatic photometry of the DAWD stars chosen as potential standards. There is a range of SEDs due to the range of $T_{e}$ and $\log g$ values derived for different DAWDs. How accurately are these reflected in the photometry? The answer to this question defines the accuracy to which DAWDs can be used as standards, as first done by \citet{Bohlin07}.  

We also immediately recognize that, since we aim to test DAWDs that are a hundred times fainter than the 3 bright CALSPEC DAWDs,  this implies that they are $\sim 10$ times farther away, at several hundred parsecs, where the effects of reddening cannot be ignored. If the reddening is modest, the colors from panchromatic photometry leave enough degrees of freedom to fit a reddening term $E(B-V)$ to resolve a discrepancy in matching the observed photometry to the predicted SEDs. The process to do so introduces several complications, and the best way to handle them drove the evolution of the analysis, as described in our series of past papers and discussed in \S\ref{sec:history}
 
\section{Project History}
\label{sec:history}

\subsection{Object Selection}
\label{sec:objselect}

We began by finding suitable candidates along the celestial equator.  DAWDs identified spectroscopically, with estimated $T_e \geq 20,000 K$ and $T_e \leq  80,000 K$ and $V \geq 16.5$ mag (with a preference for $V \sim 18$ mag)
were selected from the Villanova Catalog \citep{vill1999} and from compilations from the SDSS survey \citep[e.g.,][]{Eisenstein2006, Kleinman13}.  The equatorial objects were supplemented with similar candidates at northern declinations, also from the same sources.  
For the interested reader, further details for the selection, photometry, and spectroscopy of the equatorial and northern DAWDs were presented in C19.

At the time (before Gaia results were available), there were no equivalent faint white dwarf catalogs for southern declinations.  A list of possible candidates was compiled from objects with high proper motion and blue colors in the desired brightness range, either from lists of {\it probable} white dwarfs compiled by 
\citep[e.g.,][]{Gentile-Fusilloetal2017, Raddietal2016}, or examining (by eye) the digital images from the southern sky survey of high proper motion stars identified in the SuperCosmos Sky Survey \citep{Hambly2001}. 
Since these did not (at the time) have any confirmatory spectra in the literature,
48 candidate objects in the resulting compiled list 
were then observed spectroscopically using the Goodman spectrograph \citep{clemens2004} on the 4-meter SOAR telescope.  The spectra were obtained with signal-to-noise to permit the detailed analysis necessary if the candidates passed the final selection.  As described in C22, 19 objects were selected with an eye to keeping spatial uniformity across the southern sky.

All objects, in the celestial north, near the equator, and in the south,  were then monitored for variability using the LCO telescopes, over a range of timescales, spanning a total of 2 years. These were combined with time-domain observations available from other surveys.  Details of the process are described in \S~2 through \S~4 of C22.

The final set of 35 validated spectrophotometric standards 
is presented in Table~1 of
C22,
with the sky distribution illustrated in Fig.~1 therein.  The information is also available in Table~1 and Fig.~1 of B25.

\subsection{Photometry and Spectroscopy}
\label{sec:photspec}

The photometric data were extracted from images taken with the WFC3 camera of the HST in three separate programs.  In HST Cycle-20 (GO-12967; PI A.~Saha), 9 equatorial targets were observed in 5 passbands: $F336W$, $F475$W, $F625$W, $F775$W with the UVIS channel, and $F160$W with the IR channel. These observations were made with the full field of view (FOV) afforded by the respective detectors. 
Photometry of the targets was performed by multiple methods, as described in \S~3 of N16,
to ensure absence of systematic effects from the method used.  From the scatter in measured fluxes from multiple exposures for each target, we learned, unsurprisingly for these critically sampled images, that best results came from the images that had not undergone any resampling (i.e individual exposures that had been bias-subtracted and flat-fielded but not altered geometrically). Accuracy characterized by rms uncertainties of a few$~\times~ 0.001$ mag (the scatter from multiple exposures, as well as from the noise properties of the images themselves) were routinely attained.  

Spectra of five  of the nine targets were obtained from the Gemini Multi-Object Spectrograph (GMOS) on the Gemini South telescope.  The data are described in \S~2.2 of 
N16, and the determination of $T_{e}$ and $\log g$ from the Balmer lines is discussed in \S~4 therein. 

The subsequent analysis by 
N16,
where the photometry is compared to the synthetic photometry from the model atmosphere (using parameters derived from the spectra) showed clearly that a scaling of the standard reddening law by one parameter, ($E(B-V)$), adequately accounts for the systematic chromatic discrepancy, and leaves residuals of a few $\times 0.001$ mag 
per passband that are fully consistent with the reported photometric errors.  

These results provided a general validation for our experiment.
Nevertheless, this experiment also informed us of the need to make improvements and increase observational efficiency:

\begin{enumerate}
\item To control for systematic changes in the photometry (due to temporal drifts in the system response for whatever reason), so that all objects would be on a self consistent scale within each pass-band, we  decided to  contemporaneously observe the 3 bright DAWDs that define the CALSPEC scale.

\item Explicitly recognize the correlation between derived $T_{e}$ and reddening parameter 
$E(B-V)$.  To mitigate this, we would have to determine $T_{e}$, $\log g$ and $E(B-V)$ in one integrated process, rather than analyzing the spectra and photometry in separate sequential steps.  Additionally, we could see that adding an additional passband farther in the UV would help to resolve the degeneracy. So, an additional observation in the F275W band was added to all observations after Cycle 20 (2016). 

\item Fitting the Balmer line profiles suffers from uncertainties in knowing where the local continuum is.  The data can suffer from positioning errors at the spectrograph aperture, possibly compounded by atmospheric dispersion, all despite reasonable precautions while observing. For those interested, a detailed discussion is given in \S~2.2 of
N19.
Spectra from multiple telescopes and/or instruments help to estimate the magnitude of the problem.  Gaussian process analysis of the residuals for a trial fit of the model SED to the observed flux calibrated spectrum (see \S~6.7 of N19) 
can be used to mitigate flux calibration errors (and hence the continuum position) in the observations. 

\item 
N16
presented two different approaches to matching the model SEDs to the observed photometry.  The model parameters $T_{e}$ and $\log g$ and the resulting SEDs at the source were 
derived in the same way for both. In the first approach, the photometry was tied explicitly to the 
CALSPEC system, through identical measurements made on contemporaneous observations (e.g., F160W Program 13575; PI S. Deustua) of the three 
bright DAWD standards that define that system.  The observed colors of the candidate DAWDs were then used to compare to the model SEDs.   The second approach taken, described in \S~5 of the N19 paper, explored the potential of decoupling from the CALSPEC system and improving the accuracy attained. The essential notion is that {\it if we have self-consistent instrumental colors for all our target stars, the degree to which the instrumental colors correlate with synthetic colors (modulo fitting the reddening) is a measure of the predictability of the DAWD atmospheric models}.  The results were very encouraging and became a goal to be pursued in the longer term.

\end{enumerate}

The reader interested in probing into the details of the above issues, and how they were eventually dealt with, is referred to \S~3 of N19.

The Cycle-22 observations (program GO-13711; PI A.~Saha) observed 23 fainter DAWD targets (including repeats of those observed in Cycle-20) distributed around the celestial equator and the northern celestial hemisphere. Observations in all 6 passbands (now including $F275W$) of the 3 bright DAWDs (GD-153,  G191B2B, and GD-71) that define the CALSPEC fluxes  were interspersed over the total duration in time over which the 23 `unknowns' were observed, so that any trends in the photometric behavior of the end-to-end system could be monitored. Also, all targets in Cycle 22 and thereafter were placed in the same location of the UVIS channel (the UVIS2-C512C-SUB aperture that provides a $512 \times 512$ pixel image) in the corner of the CCD nearest the readout) to minimize charge transfer efficiency (CTE) losses.  Ground-based spectra were obtained by GMOS \citep{Hook04} on the Gemini telescopes, and also with the Blue Channel spectrograph on the MMT \citep{Angel_1979}.  Several stars were observed by both GMOS and Blue Channel, providing an opportunity to cross-check the robustness of the derived atmospheric parameters. See \S~3.7 of N19 
for details.

Cycle-25 HST observations (GO-15113; PI A.~Saha) for the southern stars utilized the same observing strategy as for Cycle-22. The process for photometry was identical as well. However, the spectroscopy of those targets was done with the Goodman spectrograph \citep{clemens2004} on the SOAR telescope. The acquisition and processing of the data are described in detail in C22.

\subsection{The Evolution of Analysis Methodology}
\label{sec:anal-meth}

Here we trace how items 1) through 4) from the list of desired improvements mentioned above in \S~\ref{sec:photspec} were progressively addressed and presented in subsequent analyses and papers.  

N19
presented results for 19 DAWDS with $16.5$ mag $ < V < 19.8 $ mag, with SEDs on the AB magnitude system.  It also comprehensively addresses items 1) through 3) from the desiderata  enumerated in \S~\ref{sec:photspec}.   Improvements to the observing program are detailed in their \S~3 and \S~4,  along with steps to account for temporal changes in sensitivity of the HST-WFC3 system and also to eliminate any systematic differences between the Cycle-20 and Cycle-22 observations (for targets observed in both cycles) that placed the primary target in different locations in the WFC3 FOV.  Once again, the photometry is tied to the CALSPEC system, using contemporaneous and identical observations of the three primary CALSPEC DAWDs. 

Section 6.7 of N19 describes how correlated errors in the flux calibration of the spectra are minimized 
(item 3 in the list from \S~\ref{sec:photspec} above).  This is followed by forward modeling the stellar parameters ($T_{e}, ~\log g, ~A_{V}$, and effective angular diameter\footnote{the relative angular diameters of the stars in the ensemble are manifest as a wavelength independent relative brightness that is solved for.}) of the DAWD in question onto the observed spectra (modulo observing conditions and spectroscopic set up particulars) and the HST multi-band photometry.  `Forward Modeling' refers to searching in the space of model parameters for the parameters that best reproduce the data. This allows examination of covariances of the parameters, as opposed to when one tries to derive the parameters directly from the data.
Details are available in their \S~6.8 through \S~6.11.   This addresses item~2 from the list of issues above. However, fluxes are still tied to the CALSPEC system as defined by Bohlin's 3 bright DAWDs, based on Hubeny's WD  \texttt{TLUSTY} models v207 and SYNSPEC53 (see \citet{Bohlin_2020} for details).  In the long run, this programmatic linkage to CALSPEC is undesirable for two reasons:
\begin{enumerate}
    \item The new candidate standards DAWDs supplemented by the 3 bright "Bohlin" DAWDs that define the CALSPEC flux scale,  can be used collectively in the same way as originally done for the 3 bright standards that define CALSPEC system as described in \citet{Bohlin_2000}, \citet{Bohlin_2003}, and
    \citet{Bohlin_2020}.
    The independence gained by doing so also provides the means to get an external estimate of uncertainties.

    \item The CALSPEC system is periodically refined (e.g., with updated models), and any asynchronicity with updating the standards presented here would result in confusion over exactly how they are interconnected. 
\end{enumerate}

These concerns were addressed in the next step of improving the process as presented by 
A23.
 This is also the first study where the full set of candidate standards were examined, including those in the southern celestial hemisphere.  Here, the final ensemble of the 35 
DAWDs follow the selection cuts described in detail in 
C22.
These are summarized 
above in \S~\ref{sec:objselect} of this paper. In the analysis, the observed and model SEDs are treated independently of the CALSPEC scale. Essentially, the modeled SEDs from atmospheric parameters and reddening models are forward modeled onto the brightnesses of each object in each of the six pass-bands, measured on a self-consistent instrumental scale (with temporal and other systematic signatures removed). The zero-points to go from the internally consistent instrumental to true colors can be determined this way for the ensemble as a whole, without reference to any external system (and allowing for the relative brightness of each object in the ensemble, through an angular diameter or distance parameter that is independent of color, one parameter for each star). Thus, a total of 4 parameters ($T_{e}, ~\log g, ~A_{V}$, and angular diameter) per star for 35 stars are varied against $6 \times 35 = 210$ measurements of brightness in the various bands, allowing for 70 degrees of freedom. 
The above process achieves the goal of making the SEDs consistent with the \citet{Hubeny95} DAWD models at the level of a few $\times 0.001$ mag per object per pass-band. The residuals are consistent with the 
independently assessed uncertainties in the photometry, and are shown graphically in Fig.~2 of A23. 
For those who may be interested, their paper discusses the impact of subtle effects (e.g., a non-standard reddening law) and looks for systematic effects in the residuals 
versus the fitted parameters.  As shown in their Fig.~6, the derived reddenings/extinctions for individual DAWDs are consistent with the 3D extinction estimates for them from \citet{Gentile-Fusillo21}, which in turn are based on a parameterization of 2D-maps of \citet{Schlegel_1998} and \citet{Schlafly11}.

While the chromatic component of the SEDs is now decoupled from CALSPEC, the achromatic component still needs an external reference. In A23, 
this was done by evaluating the average offset for the 3 bright CALSPEC DAWDs.  This amounts to one additional parameter for the whole ensemble (i.e., a common offset for all 35 DAWDs) to put them on the AB-magnitude scale. 

The A23 
analysis forward modeled the photometry, but did not forward model the spectra. $T_{e}$  and $\log g$ are allowed to vary, while being narrowly constrained about the values from those in N19.
There remained the need to combine 
the N19
analysis with the 
A23
approach to achieve simultaneous forward modeling of both spectra and photometry, while evaluating the SEDs without depending on an external reference.  Details of the process are available in \S~5 of A23, 
with results discussed in their \S~6, along with comparisons (where possible) of the model fluxes compared to other photometric systems in use, such as for PanSTARRS1, Gaia, DECam, and SLOAN, and to the 2014 version CALSPEC, which uses the same atmospheric models for DAWDS. 

The final step for full simultaneous hierarchical modeling of \emph{both} photometry and spectra while comparing internally consistent measured fluxes \emph{without} anchoring colors to an external reference like CALSPEC was achieved in B25, as was first done in A23. 
The deployment of this method achieves all of the analysis methodology goals that we had set for ourselves. The results show a slight improvement (by $\sim 0.0005$ to .001 mag) in the residuals against the WFC3 photometry.

In addition, the B25 
analysis incorporates three salient additional elements:
\begin{enumerate}
    \item 
    The DAWD model spectra used in this analysis are from \texttt{TLUSTY} v208 NLTE atmosphere models of \citep{hubeny2021}. The grid is defined from $91.2$ nm to 
    $32.0 ~\mu$m, with resolution $R=100,000$, covers $T_{e}$ from 15,000K to 70,000K, and $\log g$ from 7 to 9.5. The new models extend the grids in the infrared to 30~$\mu$m, and are normalized to a reconciliation of the absolute flux measures of Vega at 555.6 nm and Sirius in the mid IR, at 12 $\mu$m and beyond \citep{Bohlin_2020}. There is no concern about dust for Vega in the
    visible, while Sirius has no dust ring. These models are also the ones used for defining the current CALSPEC (2020) scale, so the direct comparison of our results to those from CALSPEC are more meaningful.
    \item 
    Ultraviolet STIS spectra obtained from STIS for \emph{some} of the stars became available, and the analysis methods could be applied to incorporate modeling of those data as well. 
    \item 
    The model SEDs are reddened using \citet{gordon2023}, as opposed to reddening specifications such as \citet{fitzpatrick19}, used in earlier published analyses.
\end{enumerate}

The residuals from comparing the final adopted models against the observations with HST/WFC3 are shown in Fig.~7 of B25, which is the analog to Fig.~3 of A23. The rms values have improved overall in B25 by up to $0.001$ mag.
Fig.~8 of B25
provides a visual summary of the residuals for various cases for assumed reddening behavior and/or inclusion of the ultraviolet STIS spectra, as well as a comparison with residuals from A23.

Additionally, \citet{Bohlin_2025} used the newer model grid \citep{hubeny2021} to place the 19 of our faint DAWDs that have STIS spectrophotometry explicitly on the current CALSPEC (2020) scale, using classical methods of analysis. Fig.~12 of A23
shows the comparison of the two flux scales. 
 It shows that the most noticeable change is a less than 0.01 mag shift in the overall flux, reflecting the change from the earlier Vega based nSED, to the nSED currently incorporated in CALSPEC.

As in A23,
the chromatic component of the SEDs derived for the 35 DAWDS in
B25
is independent of any external reference and stands solely on the fidelity 
of the DAWD model atmospheres. Also, as before, the 
\emph{achromatic} normalization of the SED to an absolute flux scale is determined by achromatically scaling the fluxes for \emph{all} our DAWDs by a \emph{single} common factor so that the fluxes for the 3 bright DAWDs selected by Bohlin matches their respective values in CALSPEC, as averaged over the wavelength range in question.  

In A23,
the absolute flux normalization is against an earlier version of CALSPEC, which was based on the flux for Vega from \citet{Megessier95}; and the models from \citet{Hubeny95} were used in the chromatic analysis.  In B25,
the comparison is against the current version of CALSPEC, which is on the 2020 flux scale and results in a difference in the achromatic offset by 0.0087 mag. 

B25
used an updated Hubeny WD NLTE grid computed with TLUSTY v208. The current CALSPEC (2020) is brighter than the older CALSPEC flux scale used throughout
A23.

\section{Intercomparison of Results}
\label{sec:Comps}


The results from
A23, 
\citet{Bohlin_2025}, and 
B25
provide a sense of the systematic effects and uncertainties in the results. 
We restrict this aspect of the discussion to the wavelength range covered by the WFC3 data.

The analyses by \citet{Bohlin_2025} and 
B25
utilize the same data 
and identical atmosphere models, though the grids used differed in resolution.
The analyses differ in methodology: the former is a direct analysis that references the current CALSPEC scale, and the latter a forward modeling that seeks to mitigate the effect of correlated errors, and references the same CALSPEC scale, but only for a common achromatic zero-point. The comparison of results from these two studies primarily reflect differences arising from the different analytic approaches taken. We stand by the methods developed that have led up to the B25 results, which hierarchically model using all of the available observational material, with their respective weights.


Fig.~12 in A23 
compares the derived SED for the bright DAWD GD153  using their method 
with  the \citet{Hubeny95} model and the Vega-based achromatic zero-point to the newer \citet{hubeny2021} models with the absolute flux settings based on a reconciliation with the Sirius flux scale \citep{Price2004}, as published later in \citet{Bohlin_2025}. 
The most striking feature is the 0.0087 mag over-all flux offset with smaller wavelength dependent differences.

A comparison of SEDs from 
A23
with those from 
B25, shows a weighted average achromatic difference from the three bright CALSPEC DAWDs to be $\sim 0.0117$ mag (see Table~F1 and Fig.~F1 in B25). 
 What remains after removing this difference shows
 how much the cumulative effect from upgraded models (\citet{Hubeny95} to \citet{hubeny2021}), differences in dust relations, $R_{V}$ assumptions, and enhancements to analysis methodology affect the derived SEDs. This is illustrated primarily in Fig.~B1 of B25,
which shows the differences in the modeled magnitudes from B25 versus those from A23 for all of our 35 DAWDs, for the six HST/WFC3 passbands used. 
Note that the mean difference 
stays at $\approx 0.01$ mag in all except in the near IR band {\it F160W} where it differs by $0.02$ mag, and has proportionally larger scatter, presumably from diminished accuracy from the IR detector.  The flat difference across the bands is consistent with the known achromatic difference between the two sets of SED models, which essentially explains the differences between B25 and A23 entirely. In turn, this implies that the result of the procedural improvements in B25 are subtle (within the common wavelength region where both analyses are constrained by observations). In other words, we can reasonably conclude that improvements to the analysis methods have converged. 

\section{Discussion \& Summary}
\label{sec:discussion}

\subsection{Results to Date}
\label{sec:results}
The models presented in B25 cover the wavelength range from 91.2 nm \space to 32 $\mu$m.  All 35 of the standard star models are constrained by broadband observations with HST/WFC3 between $270$ nm to $1600$ nm; 19 of them are further constrained by HST/STIS spectra. The per band, per star weighted rms difference between model and observations with HST/WFC3 is within 0.004 mag in the Near UV and visible passbands, and less than 0.007 mag in the $F160W$ passband at $1600$ nm. 

The model fluxes for each of our standard stars represent absolute calibrated fluxes at the top of the earth’s atmosphere. The synthetic model spectra were produced by TLUSTY  and SYNSPEC \citep{hubeny2021} from a grid of pure hydrogen (DA) white dwarf photospheres specified by their $T_{e}$ and $\log g$ parameters given by B25. Interstellar extinction (reddening), based on the methods of \citet{gordon2023} for the E(B-V) values in B25, has been applied to these fluxes.  Absolute flux normalization for each model is done with respect to a reconciliation by \citet{Bohlin_2025} of the absolute flux of Sirius in the mid-IR by \citet{Price2004} with that for Vega at $555.6$ nm by \citet{Megessier95}, rather than from Vega alone (this matches the normalization used for CALSPEC at the time of this writing). 

Our model SEDs for the 35 DAWDs reside in a Zenodo holding linked to B25, with each star identified by its White Dwarf Faint Standard  (WDFS) prefix and consists of tables of wavelength (in \AA) and absolute flux (ergs cm$^{-2}$ s$^{-1}$ Hz$^{-1}$).   The model spectral resolution corresponds to $R = 100,000$ and the fluxes are tabulated at a $1$~\AA ~spacing.
Thus, convolving the models to a lower spectral resolution should pose no problem.

For the foreseeable future
B25 delivers our final results in the wavelength range from $270$ nm to $1600$ nm. This range is what was covered by the WFC3 observations and includes the full set of the 35 stars. 
It is notable that the few $\times 0.001$ mag residuals per WFC3 passband per star is of the same magnitude as the uncertainties from noise in the photometry from the WFC3 images. The floor of the uncertainties of our determined SEDs clearly cannot be improved without improving the uncertainties of measurement.
It is not clear whether random measurement errors or systematic errors in both measurement and models dominate the residuals.

The final SED models, and all input data for B25 are available electronically from a Zenodo repository
linked from B25,  and directly cited as \citet{https://doi.org/10.5281/zenodo.14339960}.
Synthetic broadband magnitudes for the Dark Energy Survey (DES), Gaia,  the Panoramic Survey Telescope And Rapid Response System 1 (PS1), SDSS, and the Dark Energy Camera Legacy Survey (DECaLS) systems, along with, where available the residuals against measured magnitudes for our DAWDs on these systems  are collected in a zip file `paper\_tables.zip'
in the Zenodo repository \citet{https://doi.org/10.5281/zenodo.14339960}. 

\subsection{Qualifications}
\label{sec:SandQ}

In what follows, we highlight some qualifications, caveats, open questions, and topics for future investigation, while contrasting them against the strengths summarized above.  The procedures used to establish our White Dwarf Faint Standards provide lessons concerning how the current state of absolute and relative stellar calibrations might be improved or added to in the future.  Enhancements could include:

\begin{enumerate}
\item
The achromatic absolute flux scale remains uncertain at the ~1\% to 2\% level.  When improved laboratory based absolute fluxes become available for such bright standards as Sirius, the results can be easily applied to our SEDs by a simple multiplicative correction of that magnitude.  
\item 
The faintness of our standard stars places them at distances of hundreds of pc.  This invariably leads to varying levels of interstellar extinction, something not evident in our three brighter stars, G191B2B, GD 71, and GD 153.  For example, our faint standards exhibit $A_{v}$ values of 0.02 to 0.37 with a mean of 0.14.  This is dealt with by incorporating into our final SEDs interstellar extinction determined in terms of wavelength dependent reddening described by \citet{gordon2023} in terms of the $E(B-V)$ and $R_V$ parameters.   Clearly, it is ideal to use standard stars exhibiting minimal extinction with $E(B-V)$ values as small as possible, but to get all-sky coverage we cannot avoid reddening entirely.  We had to choose from a limited candidate list with limited resources for vetting. This situation has been completely re-vamped by Gaia, which has increased the list of suitable candidate DAWDs by orders of magnitude.  An analysis such as ours requires spectra which Gaia cannot provide, but that can come from ongoing deep all sky multi-object spectroscopic surveys such as The Dark Energy Spectroscopic Survey Instrument (DESI) on the 4m Mayall telescope in Arizona, the WHT Enhanced Area Velocity Explorer (WEAVE) on the William Hershel Telescope in La Palma and the 4-meter Multi-Object Spectroscopic Telescope (4MOST) on the Vista telescope on Cerro Paranal, Chile.   Such surveys can unambiguously identify DAWDs in the proper temperature range and flag any pathologies such as magnetic fields or other spectral peculiarities.   All of these surveys will obtain spectra of tens of thousands of DA white dwarfs, in addition to their primary cosmological targets.
\item 
We employed dedicated observing programs to place strong limits on photometric variability in our standards.  In  the future this could be accomplished from the Gaia Mission archives and from and temporal sky surveys such as the Rubin Observatory and PanSTARRS.   In summary, all the time-consuming target selection and validation observations can be satisfied in the above mentioned survey databases.
\item 
A fundamental aspect of our observations was the use of the six HST WFC3 filters to normalize our SEDs, without having to worry about extinction by the terrestrial atmosphere.  It is likely that this unique resource may not be available in the foreseeable future for when and if our observations are ever repeated or extended.  Thus, some alternative flux reference instrumentation, above the earth’s atmosphere, may be required.
\end{enumerate}

Below, we elaborate on some of these, and discuss other related topics for completeness.

\subsection{Extrapolation to other wavelengths}
\label{sec:extrap}

Our experiment constructs SEDs using models of DAWD photospheres derived from spectroscopic analysis, while simultaneously fine-tuning and normalizing to a set of six observed WFC3 filter magnitudes via synthetic photometry, and simultaneously determining the corrections for interstellar extinction. All of our stars are analyzed as an ensemble: they rely on the same analysis and systematics and are tied to the same WFC3 filters. Thus they constitute a highly self-consistent set of standard stars. The experiment also extends the validation of the proposition that modeled SEDs of hot DAWDs from the three stars by \citet{Bohlin_2000} to a much larger set of thirty-five. This validation, and the final resulting set of SEDs presented in \citet{Boyd2025} is strongest 
in the wavelength range from $275$ nm to $1600$ nm (hereafter referred to as "VIS-NIR range"), where observational constraints were provided for the \emph{full ensemble} of 35 stars from HST/WFC3 photometry fully managed within our project. In the range $\approx 91$ nm to $\approx 250$ nm (hereafter "UV range").  The observational constraint  comes from HST/STIS spectra for 19 of the 35 stars in our ensemble. These spectra help to validate our derived reddening, and the results of the UV extension that they permit are consistent with the model predictions in the VIS-NIR range. However, keeping in mind that we have not yet fully explored 
the full ensemble, and that we do not fully sample the range of possible dust reddening laws in the UV (where they express themselves the most), we hold back from endorsing the use in the UV range with as much confidence as in the VIS-NIR range. Redward of $1600$ nm, the veracity of the predicted SEDs depend almost entirely on the veracity of the DAWD atmosphere models themselves, given that $T_{e}$, $\log g$, and $A_{V}$ for all 35 stars are very well constrained by the VIS-NIR (and where applicable the UV) observations. At this time, these predicted SEDs lack observational validation red-ward of $1600$ nm, though efforts are currently underway. Nevertheless, our model SEDs, which are calculated to $32 ~\mu$m, may be the best resource for faint calibrations over  the whole sky currently available. 

\subsection{Tying SEDs to an Absolute Physical Scale} 
\label{sec:Achromaticsale}

At the time this project was begun, the driving science goals, such as for photometric redshifts, require accuracies in color of a few $\times 0.001$ mag (same as few $\times$ 0.1\%), but uncertainty in the achromatic absolute flux calibration of 1\% or 2\%  (as long as it is the same for all standards) appeared to be non-critical.  Since then, the need for characterizing the host stars of extra-solar planets places the desire for similar accuracies on the overall absolute flux. For this, our methodology of adopting the value through CALSPEC either for Vega (with an absolute flux determined at 555.6 nm  by \citet{Megessier95}),  which is beset by being a pole-on rotator and having circumstellar dust, e.g. \citet{Sibthorpe2010}; or for Sirius, where the space-based absolute flux measurement is in the mid-IR and must be projected to the visible wavelengths using its model atmosphere, are both inherently less accurate
than our determination of the relative chromatic variations in the SEDs of our standard stars. Our experiment with 35 DAWDs is a rigorous test of the SED predictions by the model atmospheres, whereas a similar test for the more complex 
atmosphere of Sirius does not yet exist, and the extrapolation from the mid-IR to the visible region is, relatively speaking, unconstrained.  The eventual solution is to wait for balloon and satellite driven laser experiments that will send measured doses of light to telescopes to achieve this calibration: see for instance, \citet{Plav2023} and \citet{Deustua2021}.  The improvements will result in an overall single achromatic adjustment to all fluxes at all wavelengths for all 35 standards from what is presented in B25.

Another possible way in the future (in principle at least) is to use the structure models for DAWDs. We can infer the radius of a DAWD from $T_{e}$ and $\log g$, along with the degenerate Mass-Radius relation $R(Mass)$, provided we know the mass.  This was demonstrated in \citet{bond_2017} for Sirius~B.  Combining such a derived radius with an accurate enough distance (from space borne parallax measurements), and hence an explicitly determined angular diameter, and also the atmospheric model derived flux density, the received flux can be used to set the achromatic zero-point.
\citet{Holberg_2012} studied twelve binary systems to verify general conformity of the WD $R(Mass)$.
However, from their analysis, the error bars at the present time 
are significantly larger than what we can achieve with the approach adopted in this work.  The uncertainties depend on the degenerate core composition, 
plus the current level of uncertainties in distances from, say, Gaia parallaxes. This precludes the use of such a method at the present time.  Note that the angular diameter need only be explicitly derived if we want to set nSEDs directly from the white dwarfs.

\subsection{SEDs and Standard Passbands}
\label{sec:nativepassbands}


The comparison of published magnitudes of our DAWDs from existing surveys (e.g., DES, Gaia, PanSTARRS, SDSS, and DECaLS) along with the synthetic magnitudes calculated from our SEDs of these WDs are in Appendix I of {B25}.
To make these comparisons, the SEDs were convolved with the published standard passbands from the respective surveys to generate the synthetic magnitudes of our  DAWDs. Offsets vary from one survey to another, but are at most $\pm 0.05$ mag.  The quoted uncertainties from the surveys concerned dominate the comparisons.

To derive the synthetic magnitudes of the DAWD standards on any native system for a given observational setup, one needs to know the effective passbands of that system. Ideally, these would be determined empirically for the full optical train: i.e. from the top of the telescope to the detector response. 


One may still resort to the color corrections used in traditional photometry to transform native photometry to standard systems. 
It is imperative to keep in mind that our DAWD standard stars are uniformly blue with SEDs that rise strongly towards short wavelengths. This should not cause any serious difficulties when calibrating spectroscopic observations. However, photometry, especially broadband photometry, of very red objects could result in biases.  It is suggested that this be investigated with numerical simulations using the particular instrumental passbands and expected SEDs of the observed objects.  Alternatively, using our sample to ‘calibrate’ red stars with known magnitudes and colors should reveal any biases.
However, used in conjunction with photometric standard stars derived from the filter convolution of their SEDs in systems like those for PanSTARRs, DECam, etc., accurate absolute colors may be derived from native photometric measurements.  There also exist images taken using HST/ACS in parallel with our HST/WFC3 observations, of adjoining fields a few arc-minutes near our DAWD standards. 
Stars in those fields will generally be much redder than the DAWDs, providing leverage for constructing color-equations as in traditional photometry. With the analysis of the DAWDs now completed, we are undertaking the reduction and publication of the photometry of suitable stars from those parallel observations. 


The above detailed discussion of limitations and caveats, which is an integral part 
of the discussion, should not detract from the unprecedented self-consistency of our results, which furnishes what the title of the paper claims.  In that vein, 
we note with a great deal of satisfaction that our results are already being used to cross-calibrate across photometric systems: \citet{popovic2025} report on how they have used the results in B25
to get improvements in the Pantheon+ and DES 5-year SN uncertainties by a factor of 1.5.  We expect that these stars and their standardized flux distributions will be a great aid for high precision astronomical measurements to come.

\acknowledgments

 We thank the anonymous referee for their careful reading of the paper, and for comments and suggestions that have helped to provide better clarity.

On behalf of the team, AS acknowledges the following NASA grants administered through STScI that specifically supported this project through its various phases:  
HST-GO-12967, HST-GO-13711, HST-GO-15113.

BMB is supported by the Cambridge Centre for Doctoral Training in Data-Intensive Science funded by the UK Science and Technology Facilities Council (STFC).

Specific acknowledgments for telescope time allocations are given the original papers where the data and their analyses are first described. Here we credit the various facilities used.

\facilities{\emph{HST} (WFC3, STIS), \emph{SOAR}, \emph{MMT}, \emph {Gemini}, \emph{Gaia}, \emph{PaNSTARRS}, \emph{SDSS}, \emph{DES}, \emph{DECaLS},
 \emph{Las Cumbres Observatory}.
}
\vskip5mm

\bibliographystyle{aasjournal}
\bibliography{references}

\end{document}